\documentclass[10pt,conference]{IEEEtran}
\IEEEoverridecommandlockouts
\usepackage{cite}
\usepackage{amsmath,amssymb,amsfonts}
\usepackage{algorithmic}
\usepackage{graphicx}
\usepackage{textcomp}
\usepackage{xcolor}
\usepackage{todonotes}
\usepackage[normalem]{ulem}
\usepackage{ifthen}
\usepackage{xspace}
\usepackage{soul}
\usepackage{flushend}
\usepackage[utf8]{inputenc}
\usepackage[english]{babel}
\usepackage{tabularx}
\usepackage{booktabs}
\usepackage{latexsym}
\usepackage{mathrsfs}
\usepackage{enumerate}
\usepackage{colortbl}
\usepackage{longtable}
\usepackage{tabu}
\usepackage{balance}
\usepackage{multirow}
\usepackage{epstopdf}
\usepackage{paralist}
\usepackage{adjustbox}
\usepackage{color, colortbl}
\usepackage[inline]{enumitem}
\usepackage{filecontents}
\usepackage{listings}
\usepackage{tcolorbox}
\usepackage{arydshln}
\usepackage{nccmath}
\usepackage{lipsum}
\usepackage{pdfcomment}
\usepackage{url}
\usepackage{listings}
\usepackage{tcolorbox}
\usepackage{cleveref}[noabbrev]

\usepackage{adjustbox}
\usepackage[T1]{fontenc}
\usepackage{upquote}
\usepackage[caption=false]{subfig}

\usepackage{cancel}

\usepackage{hyperref}
\hypersetup{
     colorlinks   = true,
     citecolor    = blue,
     linkcolor    = blue,
     urlcolor=blue
}

\def\BibTeX{{\rm B\kern-.05em{\sc i\kern-.025em b}\kern-.08em
    T\kern-.1667em\lower.7ex\hbox{E}\kern-.125emX}}
\begin{document}

\title{Why Don’t Developers Detect Improper Input Validation? \texttt{'; DROP TABLE Papers; -\xspace-}}

\author{\IEEEauthorblockN{Larissa Braz}
	\IEEEauthorblockA{\textit{University of Zurich}\\
		larissa@ifi.uzh.ch}
	\and
	\IEEEauthorblockN{Enrico Fregnan}
	\IEEEauthorblockA{\textit{University of Zurich} \\
			fregnan@ifi.uzh.ch}
		\and
		\IEEEauthorblockN{Gül Çalikli}
		\IEEEauthorblockA{\textit{University of Zurich} \\
				calikli@ifi.uzh.ch}
			\and
			\IEEEauthorblockN{Alberto Bacchelli}
			\IEEEauthorblockA{\textit{University of Zurich} \\
					bacchelli@ifi.uzh.ch}
}

\maketitle

% !TEX root = ../security-developers.tex

\newcommand{\etal}{\textit{et al.}\xspace}
\newcommand{\eg}{\textit{e.g.,}\xspace}
\newcommand{\ie}{\textit{i.e.,}\xspace}

\newcommand{\sqliShort}{\textbf{SQLI}\xspace}
\newcommand{\iiv}{\textit{Improper Input Validation}\xspace}
\newcommand{\sql}{\textit{SQL Injection}\xspace}
\newcommand{\iivShort}{\textbf{IIV}\xspace}
\newcommand{\ivqiShort}{\textbf{IVQI}\xspace} 
\newcommand{\ivqi}{\textit{Improper Validation of Specified Quantity Input}\xspace}

\newcommand{\perparticipantsFoundBug}{\textcolor{red}{XX\%}\xspace}
\newcommand{\perparticipantsFoundBugANDvuln}{\textcolor{red}{XX\%}\xspace}
\newcommand{\numparticipantsFoundBugNOTvuln}{\textcolor{red}{XX}\xspace}
\newcommand{\authorWhoAnalysedCRanswers}{second\xspace}
\newcommand{\authorWhoDidQualityCheckforCRanswers}{first\xspace}

\newcommand{\numOfSSEsInterview}{four\xspace}

\newcommand{\numOfTotalAccess}{472\xspace}

\newcommand{\figshare}{\url{https://doi.org/10.5281/zenodo.3996696}}

%--- start ---- Research Questions -------------
\newcommand{\rqOne}{Do developers detect Improper Input Validation (\iivShort) vulnerabilities during code review?\xspace}
\newcommand{\rqOneOne}{To what extent do developers detect \iivShort vulnerabilities during code review?\xspace}
\newcommand{\rqOneTwo}{What is the effect of the visibility of a traditional attack scenario for an \iivShort vulnerability on its detection during code review?\xspace}

\newcommand{\rqTwo}{What is the effect of warning developers who missed the \iivShort
	about the existence of a vulnerability (i.e., prompting) on the detection of an \iivShort?}

\newcommand{\HNullOne}{The visibility of a traditional attack scenario for an \iivShort vulnerability does not affect its detection during code review.\xspace}
\newcommand{\HNullTwo}{Prompting does not affect the detection of an \iivShort vulnerability during code review.\xspace}
%--- end ----  Research Questions -------------

%--- start ----Logistic Regression Variables-------------
\newcommand{\vf}{VulnFound\xspace} 
\newcommand{\vnf}{VulnNotFound\xspace} 
\newcommand{\vfl}{VulnFound2\xspace} 
\newcommand{\vnfl}{VulnNotFoundLater\xspace}
\newcommand{\vulntype}{VulnType\xspace} 
\newcommand{\familiar}{Familiarity\xspace}  

% --------Variables that contribute to the factor "Security Practice"
\newcommand{\incidents}{Incidents\xspace} 
\newcommand{\vulexp}{VulnExp\xspace} 
\newcommand{\intvuln}{IntroducedVuln\xspace}
\newcommand{\foundvuln}{FoundVuln\xspace} 
\newcommand{\fixvuln}{FixedVuln\xspace}
\newcommand{\exploitvuln}{ExploitedVuln\xspace}
\newcommand{\practice}{UseInPractice\xspace} 

% --------Variables that contribute to the factor "Security Training"
\newcommand{\courses}{Courses\xspace} 
\newcommand{\training}{TrainingType\xspace} 
\newcommand{\lec}{Lectures\xspace}  
\newcommand{\conf}{Conferences\xspace}  
\newcommand{\sem}{Seminars\xspace}  
\newcommand{\handson}{HandsOn\xspace}  
\newcommand{\proftrain}{ProfTraining\xspace} 
\newcommand{\update}{KnowledgeUpdate\xspace}  

% --------Variables that contribute to the factor "Code Review vs. Tool Usage"
\newcommand{\static}{StaticAnalysis\xspace} 
\newcommand{\dynamic}{DynamicAnalysis\xspace} 
\newcommand{\manual}{ManualAnalysis\xspace}

% --------Variables that contribute to the factor "Mindset"
\newcommand{\responsible}{Responsibility\xspace} 
\newcommand{\bug}{BugFound\xspace}  

% --------Variables that contribute to factors "Mindset" and "Security Practice"
\newcommand{\design}{Design\xspace}
\newcommand{\code}{Coding\xspace}
\newcommand{\review}{Reviewing\xspace} 

% --------Variables that contribute to the factor "Team Issues"  
\newcommand{\aware}{Awareness\xspace} 
\newcommand{\expert}{Expertise\xspace} 
\newcommand{\tools}{ToolUsage\xspace} 
\newcommand{\library}{ThirdPartyLib\xspace} 
\newcommand{\CRuse}{CRusage\xspace} 
\newcommand{\timepressure}{EnoughTime\xspace} 

% --------Variables that contribute to factors "Security Training" and "Team Issues"  
\newcommand{\docTr}{DocAndTraining\xspace}

%Control Variables
\newcommand{\gender}{Gender\xspace}
\newcommand{\loe}{LevelOfEducation\xspace}
\newcommand{\emp}{EmploymentStatus\xspace}
\newcommand{\role}{Role\xspace} 
\newcommand{\companysize}{CompanySize\xspace}
\newcommand{\teamsize}{TeamSize\xspace}
\newcommand{\oss}{OSSDev\xspace} 
\newcommand{\pde}{ProfDevExp\xspace} 
\newcommand{\je}{JavaExp\xspace}
\newcommand{\rp}{ReviewPractice\xspace}
\newcommand{\re}{ReviewExp\xspace}
\newcommand{\web}{WebDevExp\xspace}
\newcommand{\db}{DBDevExp\xspace}
\newcommand{\designFreq}{DesignFreq\xspace}
\newcommand{\codingFreq}{DevFreq\xspace}
\newcommand{\crFreq}{CRFreq\xspace}
\newcommand{\inter}{Interruptions\xspace}
\newcommand{\interfirst}{InterruptionsFirst\xspace}
\newcommand{\intersecond}{InterruptionsNext\xspace}
\newcommand{\tdexp}{DurationExp\xspace}
\newcommand{\tdrevisit}{DurationRevisit\xspace}
\newcommand{\reviewDuration}{DurationReview\xspace}

\newcommand{\pvalue}{\emph{p}\xspace}

\definecolor{gray50}{gray}{.5}
\definecolor{gray40}{gray}{.6}
\definecolor{gray30}{gray}{.7}
\definecolor{gray20}{gray}{.8}
\definecolor{gray10}{gray}{.9}
\definecolor{gray05}{gray}{.95}

\newlength\Linewidth
\def\findlength{\setlength\Linewidth\linewidth
	\addtolength\Linewidth{-4\fboxrule}
	\addtolength\Linewidth{-3\fboxsep}
}

\newenvironment{rqbox}{\par\begingroup
	\setlength{\fboxsep}{5pt}\findlength
	\setbox0=\vbox\bgroup\noindent
	\hsize=0.95\linewidth
	\begin{minipage}{0.95\linewidth}\normalsize}
	{\end{minipage}\egroup
	\textcolor{gray20}{\fboxsep1.5pt\fbox
		{\fboxsep5pt\colorbox{gray05}{\normalcolor\box0}}}
	\endgroup\par\noindent
	\normalcolor\ignorespacesafterend}
\let\Examplebox\examplebox
\let\endExamplebox\endexamplebox

\newcommand{\rb}[1]{
	
	\vspace{0.3cm}
	\begin{tcolorbox}[colback=gray!05,
		colframe=black,
		width=\columnwidth,
		arc=3mm, auto outer arc,
		boxrule=0.5pt,
		]
		#1
	\end{tcolorbox}
}

\newcounter{Finding}
\stepcounter{Finding}

\newcommand{\roundedbox}[1]{
	\rb{
		\noindent
		\textit{\textbf{Finding \theFinding}. #1}
	}
	\stepcounter{Finding}
}

\newcommand{\numParticipantsExperiment}{146\xspace}
\newcommand{\numParticipantsMale}{109\xspace}
\newcommand{\numParticipantsFemale}{7\xspace}
\newcommand{\numParticipantsGenderNotdisclosed}{30\xspace}

\newcommand{\numParticipantsSQL}{80\xspace}
\newcommand{\numParticipantsIIV}{66\xspace}

\newcommand{\numParticipantsSQLfoundFirstCR}{52\xspace}
\newcommand{\numParticipantsSQLfoundSecondCR}{13\xspace}
\newcommand{\perParticipantsSQLfoundFirstCR}{65\%\xspace}
\newcommand{\perParticipantsSQLfoundSecondCR}{16\%\xspace}
\newcommand{\perParticipantsSQLfoundSecondCRHadMissedFirstCR}{46\%\xspace}

\newcommand{\numParticipantsIIVfoundFirstCR}{14\xspace}
\newcommand{\numParticipantsIIVfoundSecondCR}{7\xspace}
\newcommand{\perParticipantsIIVfoundFirstCR}{18\%\xspace}
\newcommand{\perParticipantsIIVfoundSecondCR}{9\%\xspace}

\newcommand{\numParticipantsSQLfound}{65\xspace}
\newcommand{\perParticipantsSQLfound}{81\%\xspace}

\newcommand{\numParticipantsIIVfound}{21\xspace}
\newcommand{\perParticipantsIIVfound}{32\%\xspace}

\newcommand{\numParticipantsFound}{86\xspace}
\newcommand{\perParticipantsFound}{59\%\xspace}

\newcommand{\numParticipantsMissedCRfirst}{80\xspace}
\newcommand{\numParticipantsMissed}{60\xspace}
\newcommand{\perParticipantsMissed}{41\%\xspace}

\newcommand{\numParticipantsSQLmissedCRfirst}{28\xspace}
\newcommand{\numParticipantsSQLmissed}{15\xspace}
\newcommand{\perParticipantsSQLmissed}{19\%\xspace}

\newcommand{\numParticipantsIIVmissed}{45\xspace}
\newcommand{\perParticipantsIIVmissed}{68\%\xspace}
\newcommand{\perParticipantsIIVmissedRateAllMissed}{91\%\xspace}

\newcommand{\numParticipantsFoundSecondCR}{20\xspace}
\newcommand{\perParticipantsFoundSecondCR}{14\%\xspace}

\newcommand{\numParticipantsFoundFirstCR}{66\xspace}
\newcommand{\perParticipantsFoundFirstCR}{45\%\xspace}

\newcommand{\numParticipantsSD}{83\xspace}
\newcommand{\perParticipantsSD}{57\%\xspace}

\newcommand{\numParticipantsSDMoreThree}{105\xspace}
\newcommand{\perParticipantsSDthreeyears}{23\%\xspace}
\newcommand{\perParticipantsSDsixyears}{32\%\xspace}
\newcommand{\perParticipantsSDelevenyears}{18\%\xspace}
\newcommand{\perMoreThanSixExperience}{50\%\xspace}

\newcommand{\perParticipantsSDdesignDaily}{61\%\xspace}
\newcommand{\perParticipantsSDprogramDaily}{96\%\xspace}
\newcommand{\perParticipantsSDcrDaily}{64\%\xspace}

\newcommand{\oddsRatioshort}{7\xspace}
\newcommand{\oddsRatio}{6.90 (3.27,14.57)\xspace}
\newcommand{\oddsRatioshortTXT}{seven\xspace}
\newcommand{\rhofirst}{$<$  0.001\xspace}

\newcommand{\oddsRatioshortSecond}{6\xspace}
\newcommand{\oddsRatioSecond}{5.57 (1.88,16.55)\xspace}
\newcommand{\oddsRatioshortSecondTXT}{six\xspace}
\newcommand{\rhosecond}{$<$ 0.001\xspace}

\newcommand{\oddsRatioshortComparing}{2\xspace}
\newcommand{\oddsRatioComparing}{2.00 (0.67,5.96)\xspace}
\newcommand{\oddsRatioshortComparingTXT}{two\xspace}
\newcommand{\numParticipantsFoundBug}{76\xspace}
\newcommand{\numParticipantsSQLFoundBug}{39\xspace}
\newcommand{\perParticipantsSQLFoundBug}{49\%\xspace}
\newcommand{\numParticipantsIIVFoundBug}{37\%\xspace}
\newcommand{\perParticipantsIIVFoundBug}{56\%\xspace}
\newcommand{\rhoComparing}{= 0.21\xspace}
\newcommand{\numParticipantsCRsecond}{80\xspace}
\newcommand{\numParticipantsIIVmissedcrfirst}{80\xspace}

\begin{abstract}
Improper Input Validation (IIV) is a software vulnerability that occurs when a system does not safely handle input data. 
Even though \iivShort is easy to detect and fix, it still commonly happens in practice.

In this paper, we study to what extent developers can detect \iivShort and investigate underlying reasons.
This knowledge is essential to better understand how to support developers in creating secure software systems.
We conduct an online experiment with \numParticipantsExperiment participants, of which \numParticipantsSDMoreThree report at least three years of professional software development experience. %comprising a code review task, a survey, and a pretest-posttest study. 
Our results show that the existence of a visible attack scenario facilitates the detection of \iivShort vulnerabilities and that a significant portion of developers who did not find the vulnerability initially could identify it when warned about its existence. Yet, a total of \numParticipantsMissed participants could not detect the vulnerability even after the warning. 
Other factors, such as the frequency with which the participants perform code reviews, influence the detection of \iivShort.
Data and materials: \figshare

\end{abstract}
% !TEX root = ../security-developers.tex

\section{Introduction}
\label{sec:intro}

A software vulnerability is ``a design flaw or an implementation bug that allows an attacker to cause harm to the stakeholders of an application''~\cite{owaspDefinition}.
To avoid vulnerabilities, much effort has been spent on making web applications more secure~\cite{Dhamankar:2009,CWETop25,OwaspTopTen}, and organizations are shifting security to earlier stages of software development, such as during code review~\cite{gitlab-survey}. Nevertheless, dangerous vulnerabilities are still routinely discovered~\cite{cve-details}.

One of the most prevalent and high-risk vulnerabilities to this day is \iiv (\iivShort)~\cite{CWETop25,tsipenyuk2005seven}. This vulnerability  is the root cause of more than half of the top ten vulnerabilities in the CWE Top 25 list~\cite{CWETop25} and is present when a software system does not ensure that the received input can be processed safely and correctly.

From a software engineering standpoint, an interesting peculiarity of \iivShort is that it does not require hard-core security expertise to be caught and avoided~\cite{owaspCheatSheet}. So, \emph{why do developers not recognize \iiv}? Answering this is central to understand how to support developers in building more secure software systems.

One possible answer is that most software developers do not possess even a very basic knowledge of software security: Several surveyed security experts believe that less than half of the developers can spot security holes~\cite{gitlab-survey}. Another possible answer is the lack of a proper attitude towards security in information systems professionals~\cite{Woon:2007}. Xie et al.~\cite{Xie:2011} suggest that software developers have general knowledge and awareness of software security and point out the gap between developers' security knowledge and their behaviors. According to the authors' findings, 
developers do not perform secure development due to factors related to their mindset, such as relying on third parties (\eg security experts), and other phases of the development life cycle (\eg design phase), besides external constraints (\eg deadlines, budget, customer demands, regulations). 
These studies report the \emph{perception} of developers and security engineers, which may or may not reflect the real situation. Further studies are needed to verify and complement these claims.

In this paper, we present the design and execution of an online experiment we devised to investigate to what extent developers can(not) detect \iivShort and the underlying reasons. We hypothesize that the visibility of a traditional attack scenario (specifically, a \sql pattern) affects whether developers can detect an \iivShort. Attack scenarios for \sql (\sqliShort) are readily available even in textbooks~\cite{Clarke:2012, Galluccio:2020} and popular culture~\cite{xkcd}, whereas for the detection of other \iivShort vulnerabilities, developers need to discover attack scenarios and assess their possibility to occur themselves. Based on the aforementioned previous studies~\cite{Xie:2011, Woon:2007}, we also investigate the effect of informing developers about the existence of a vulnerability (i.e., prompting) on its detection. 
Inspired by previous work~\cite{Naiakshina:2017, Naiakshina:2018, Naiakshina:2019} on how prompting facilitates secure password storage, we hypothesize that some developers have the knowledge to find an \iivShort but need to be prompted to focus on it. Finally, we also investigate what the developers report as the reasons why they found or not these vulnerabilities.  

We set up our experiment as an online study composed of a code review task, a survey, and a repeated review after prompting. 
We received valid responses from \numParticipantsExperiment participants, 82.5\% of whom report to be software developers, and 105 have three or more years of professional programming experience. Our main findings include: (1) The visibility of an attack scenario greatly facilitates the detection of \iivShort;
(2) prompting has an effect on \iivShort detection, yet many participants cannot detect an \iivShort without a traditional attack scenario; (3) security awareness during code development and frequency of code review play a role in the detection of \iivShort.

Based on our findings, we discuss implications and outline future avenues for research and practice.

% !TEX root = ../security-developers.tex

\section{Related Work}
\label{sec:related}

The study of software vulnerabilities and their developer-related factors is a growing area of research in the Software Engineering (SE) research community (\eg~\cite{morrison2018identifying,santos2019empirical,rahman2019seven,mirakhorli2020understanding}).\smallskip

\noindent\textbf{Improper Input Validation.}
Among vulnerabilities, {\iivShort}s are prevalent and well-known.
Scholte \etal~\cite{Scholte:2012} investigated the evolution of \iivShort vulnerabilities over a decade with a specific focus on $\sqliShort$ and cross-site scripting. The authors found that the \emph{attack surface} of \sqliShort is much smaller than for cross-site scripting. The attack surface is the set of points on the boundary of a system or an environment where an attacker can try to enter, cause an effect on, or extract data from it \cite{Howard:2019}. Our study aims to investigate the extent to which developers detect \iivShort vulnerabilities. Therefore, we focus on \sqliShort, which has a relatively small attack surface, hence is easier to detect, and another \iivShort vulnerability, namely \textit{Improper Input Validation for specified Quantity Input} (\ivqiShort), which also has a small attack surface but does not have such a visible attack scenario.\smallskip

\noindent\textbf{Why developers do not spot vulnerabilities.}
Regarding the underlying causes of why developers introduce and cannot recognize security issues, the study by Woon and Kankanhalli~\cite{Woon:2007} is based on the claim that Information Systems (IS) professionals' intention to practice secure development, which somehow relates to their mindset, is a primary cause for vulnerabilities. The authors investigated factors that are likely to impact IS professionals' intention to secure development through a survey conducted with 184 IS professionals.  
They found that attitude (determined by the usefulness of security practices to the final software product and IS professionals' career as well as subjective norms) significantly impacts the intention to practice secure development. 
Xie \etal~\cite{Xie:2011} reported gaps between developers' knowledge and the actual practices and behaviors as underlying causes for software vulnerabilities.  Xie \etal point out the developers' mindset as a main underlying cause for software vulnerabilities. We also focus on this aspect and investigate whether warning developers about the existence of a vulnerability in the code enable its detection. Findings by Xie \etal~\cite{Xie:2011} are based on semi-structured interviews that authors conducted with 15 developers, a relatively small sample size. We conducted our study with 146 participants, 57\% of whom currently work as a developer, and overall 71\% of them work as software practitioners. Our study collects and investigates qualitative and quantitative data while participants are in action during code review through an experiment and as data reflecting participants' perspectives that are collected through surveys.\smallskip

\noindent\textbf{Code Review for Vulnerability Detection.}
Previous research~\cite{Meneely:2012, Shin:2011, Meneely:2010} found that more code reviews and reviewers have positive effects on secure software development, whereas the results on the Chromium project by Meneely \etal~\cite{Meneely:2014} contradict these findings. In a study with more than three thousand Open Source Software (OSS) projects, Thompson and Wagner~\cite{Thompson:2017} concluded that code review reduces security bugs. Another study on OSS projects~\cite{bosu:2014} found that code review can identify common types of vulnerabilities. On the other hand, Edmundson \etal~\cite{Edmundson:2013} conducted a study to investigate the effectiveness of a security-focused manual code review of web applications containing SQL injections. The authors' results indicate relatively low effectiveness in vulnerability detection. In line with these results, di Biase \etal~\cite{Biase:2016} found that approximately only 1\% of Chromium's review comments are about potential security flaws. While these studies ~\cite{Thompson:2017, Edmundson:2013,bosu:2014, Biase:2016} mainly focus on the effectiveness of code review on vulnerability detection, our study investigates whether developers can(not) detect vulnerabilities (specifically \iivShort) during code review and the underlying causes with a focus on developers' knowledge and mindset.

% !TEX root = ../security-developers.tex

\section{Methodology}
\label{sec:metholdolgy}

\begin{figure*}[h]
	\centering
	\includegraphics[width=1\textwidth]{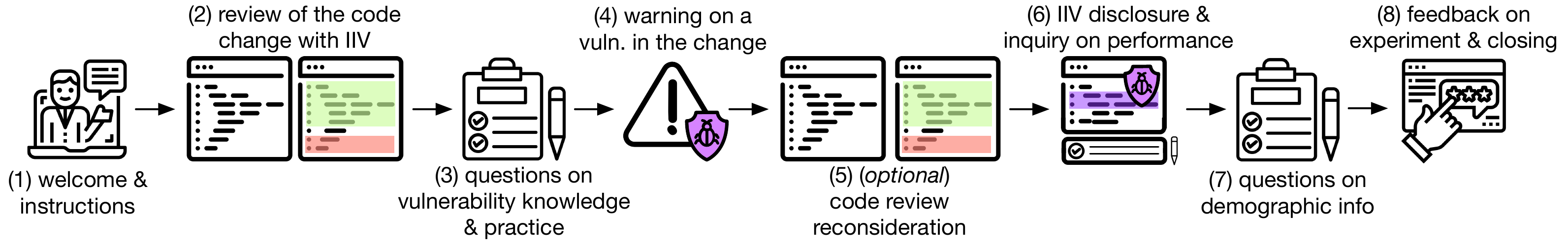}
	\caption{Design and flow of the online experiment.} 
	\label{fig:flow}
\end{figure*}

Overall, our research aims to understand to what extent developers can(not) detect \iivShort vulnerabilities and why. We base our study on an online experiment with several steps (summarized by \Cref{fig:flow}) that we devised to collect different types of evidence as well as self-reported data.

\subsection{Research Questions}
\label{sec:design:rqs}

We structured our study in two main research questions. With the first research question, we investigate the extent to which developers can detect an \iivShort vulnerability as well as the effect of the visibility of a traditional (\ie textbook) attack scenario for an \iivShort vulnerability on its detection.

\begin{center}
	\begin{rqbox}
		\begin{description}	    
	               \item[]\textbf{RQ$_1$.} \emph{\rqOne}
		 \end{description} 
	\end{rqbox}
\end{center} 

\noindent We organize our research question as follows: 

\begin{description}
	\item[\textbf{RQ$_{1.1}$.}]  \rqOneOne
	\item[\textbf{RQ$_{1.2}$.}]  \rqOneTwo
\end{description}

\noindent In particular, we test the following hypothesis for \textbf{RQ$_{1.2}$}:

\begin{description}[leftmargin=0.3cm]
	\item[\textbf{H0$_{1}$:}]  \HNullOne	
\end{description}

Woon and Kankanhalli \cite{Woon:2007} and Xie et al. \cite{Xie:2011} argued that the mindset might be the reason why developers do (not) detect vulnerabilities: Developers may not pay attention to vulnerabilities because it is not their normal role/practice. We test this hypothesis for \iivShort vulnerabilities. To do so, we take inspiration from the studies by Naiakshina et al.~\cite{Naiakshina:2017, Naiakshina:2018, Naiakshina:2019}, who investigated the effects of prompting on the implementation of secure password storage. After participants complete their first review and answer questions about vulnerabilities, we warn them about the existence of a vulnerability in the code they just reviewed and ask them to reconsider their review if they missed it. This way, we investigate whether warning developers (who missed the \iivShort) about a vulnerability's existence affects their ability to detect the \iivShort. Therefore, our second research question is: 

\begin{center}
	\begin{rqbox}
		\begin{description}	    
	               \item[]\textbf{RQ$_2$.} \emph{\rqTwo}
		 \end{description} 
	\end{rqbox}
\end{center} 

\noindent In particular, we test the following hypothesis:

\begin{description}[leftmargin=0.3cm]
	\item[\textbf{H0$_{2}$:}] \HNullTwo
\end{description}

\subsection{Design Overview}
\label{sec:design}
Our study is implemented as an online experiment that can be reached through a public website. In the following, we detail our study's design and how the experiment flows through each step. Each step corresponds to a different webpage, and returning to previous pages is not allowed.\smallskip

\noindent\textbf{(1) Welcome page:} On the first page of the experiment, we provide participants with information about the study. We introduce ourselves as researchers investigating ways to improve code review. We do not inform the participants about the study's final focus on software vulnerabilities to avoid that they form an interpretation of the purpose of the study and subconsciously change their behavior to fit it (\ie demand characteristics~\cite{Nichols:2008}).
	We also inform participants about data handling policy and ask for their consent to use their data.

\noindent\textbf{(2) Code review task:} In this step, each participant is asked to perform a code review of a code change. The code change includes a method that does not validate the input received from its external users (i.e., \iivShort). There are two slightly different versions (treatments) of this change, and each participant is randomly assigned to either one:

\begin{description}[leftmargin=0.3cm]
		\item[\sqliShort:] The received input is directly used as part of a SQL query making the code vulnerable to \sql~\cite{CWE-89}. The construction of the SQL query is visible in the code, thus showing a traditional attack scenario.
		
		\item[\ivqiShort:] The received input includes an integer amount used without validating its boundaries, thus making the code vulnerable to an \ivqi~\cite{CWE-1284}. Particularly, a negative integer value benefits the users in an unwanted way.

\end{description}
	The changed code visible in the review is sufficient to detect both vulnerabilities and how they can be exploited. Both vulnerabilities belong to ``Improper Input Validation (\iivShort)''~\cite{CWE-20}.
	
	In addition to the vulnerability (i.e., either \sqliShort or \ivqiShort), we introduce an algorithmic bug (a corner case, CC, bug) in the code change. We use this bug as a robustness check to analyze participants' interest in the task. 
	Moreover, immediately after the code review, we ask the participants whether they were interrupted during the review and, if so, for how long.
	
\noindent\textbf{(3) Knowledge and practice regarding vulnerabilities:} 
	In this step, we let the participants know that the study is about security and provide the definition of software vulnerability~\cite{Owasp}. 
	We ask questions to gather information about factors that may affect the participants' ability to detect the \iivShort vulnerability (\sqliShort or \ivqiShort), such as their security knowledge, practices, and team culture. Most of the questions are closed in a Likert scale format (the exact questions are available in the accompanying material~\cite{replication-package}).
		
\noindent\textbf{(4) Warning about a vulnerability in the change:} In this step, we notify the participants that the change they just reviewed has a vulnerability. We do not specify the type of vulnerability to avoid making its discovery too straightforward since we want to focus on the mindset's shift.

\noindent\textbf{(5) Code review reconsideration (optional):} Participants are asked to re-inspect the code review unless they think they already found the vulnerability during the first review (hence this step is optional). In the end, we ask participants if they were interrupted during this review and, if so, for how long.
	
\noindent\textbf{(6) \iivShort disclosure and inquiry on performance:} We show the type and location of the \iivShort vulnerability that affects the change they reviewed and explain how it makes the code vulnerable.
Then, we ask the participants whether they found the vulnerability. If not, we ask them to explain why they missed it. If so, we ask in which review they found the vulnerability: If in the first one, we ask them to explain why they could catch it, if in the second one, we ask why they think they could find it only in the second trial. With this step, we aim to check whether the developers agree with our 'miss/found' evaluation of the vulnerability and collect rich qualitative data to triangulate our findings and answer our RQs.

\noindent\textbf{(7) Demographics:} Participants are asked to fill in questions to collect demographic information and confounding factors, such as gender, highest obtained education, years of experience (all the questions are available in the accompanying material~\cite{replication-package}).
This information is mandatory to fill in since collecting such data helps us identify which portion of the developer population is represented by our study participants~\cite{Falessi:2018}.

\noindent\textbf{(8) Feedback and closing:} In the end, we ask the participants for feedback on the overall study. We also ask participants if they would like to share their data anonymously in a public research dataset and receive the study results.

\subsection{Design Implementation}
To implement our design, we extend the publicly available browser-based tool CRExperiment~\cite{crexperiment}. CRExperiment is designed to conduct online experiments that require participants to review code changes and answer survey-like questions; it has been used and validated in previous studies~\cite{spadini2019test,spadini2020primers}. CRExperiment uses Mergely~\cite{Mergely}, which is also used by the popular review tool Gerrit~\cite{Gerrit}, to show code changes in two-pane diffs. The Graphical User Interface (GUI) has the same color scheme as Gerrit to facilitate the simulation of a real-world code review scenario during the experiment. In addition to the answers we collect through explicit questions and tasks, CRExperiment also logs user interactions (\eg mouse clicks and pressed keys), which we use to ensure that participants actively perform the tasks. Finally, CRExperiment logs the time participants spend at each stage of the study. We store all the collected data on a server anonymously. Finally, to reduce the risk of data loss and corruption, we store the data in its raw form (i.e., recorded as logs) for offline analysis.

\subsection{Experimental Objects}

The experiment objects are a code change to review and an \iivShort vulnerability (either \sqliShort or \ivqiShort) we injected in the code change. We also inject a control bug to use as a robustness check. All the material is available in our replication package~\cite{replication-package}.

\noindent\textbf{Code Change.} Our requirements for designing the object code change are: (1) written in Java, one of the most popular languages~\cite{tiobe}; (2) not belonging to any existing code base to avoid giving some developers an advantage over the others due to familiarity with code; (3) suitable for the injection of both vulnerabilities (i.e., \sqliShort and \ivqiShort); (4) suitable to be part of an actual software (i.e., not a toy example available on websites aiming to teach beginners Java programming~\cite{Programiz}); (5) self-contained.
After several brainstorming sessions among the authors, the first version of the patch was implemented. Based on the feedback we received from the pilot studies (\Cref{sec:pilot}), we iteratively modified the patch. The final version was discussed and evaluated with two senior software developers with more than ten years of professional software development experience with large software companies. This step led to the last modifications that ensured the change did not have any implementation or design-related issues other than the vulnerabilities and the bug. The change implements a feature to manage employees' vacations, modifying two classes and six methods. The change with \sqliShort has 137 lines of code, while the change with the \ivqiShort has 145. \smallskip

\begin{figure}[t]%
	\centering
	\subfloat[\sql (\sqliShort)]{%
		\includegraphics[clip,width=\columnwidth]{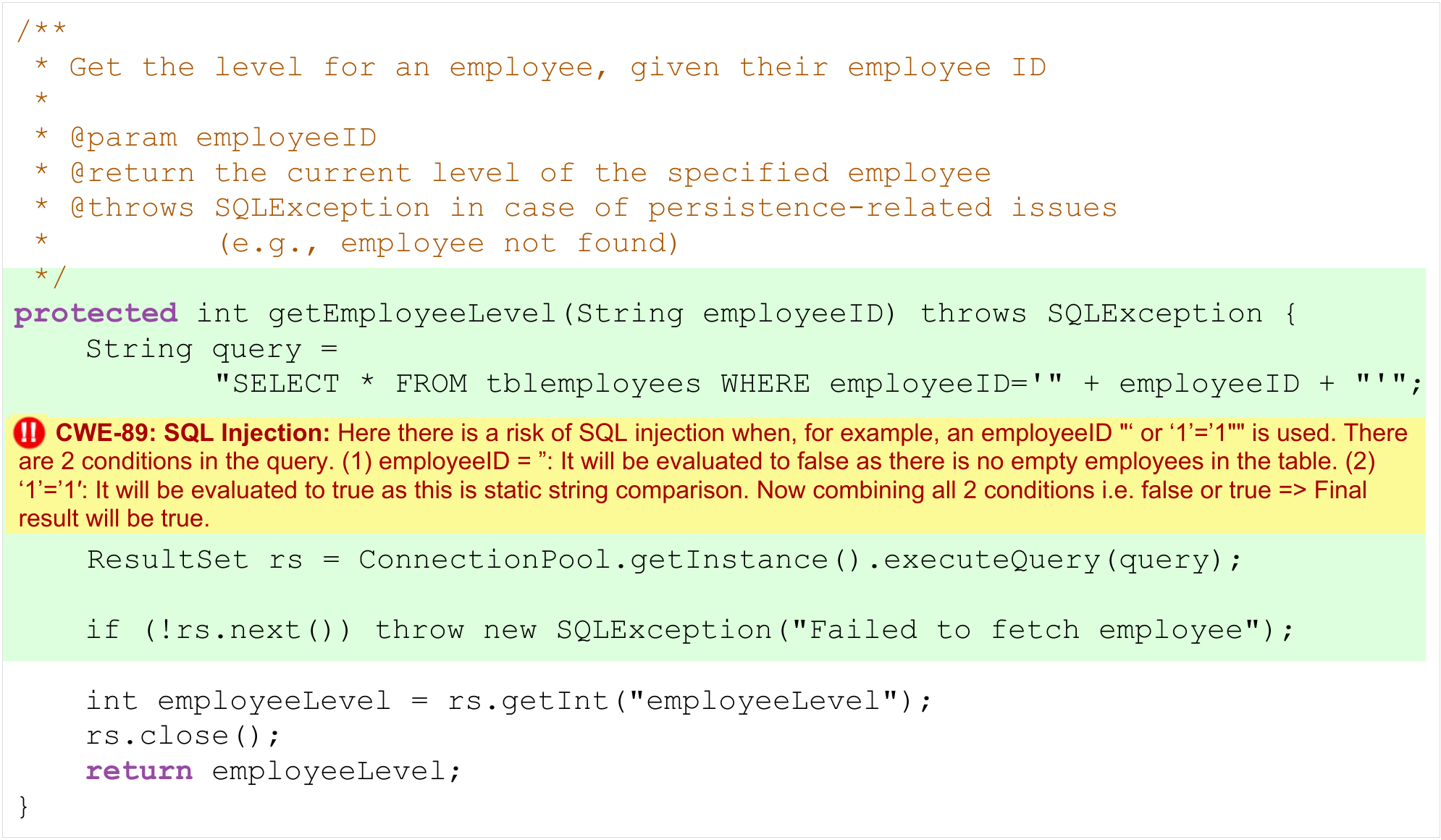}%%
	}
	
	\subfloat[\ivqi (\ivqiShort)]{%
		\includegraphics[clip,width=\columnwidth]{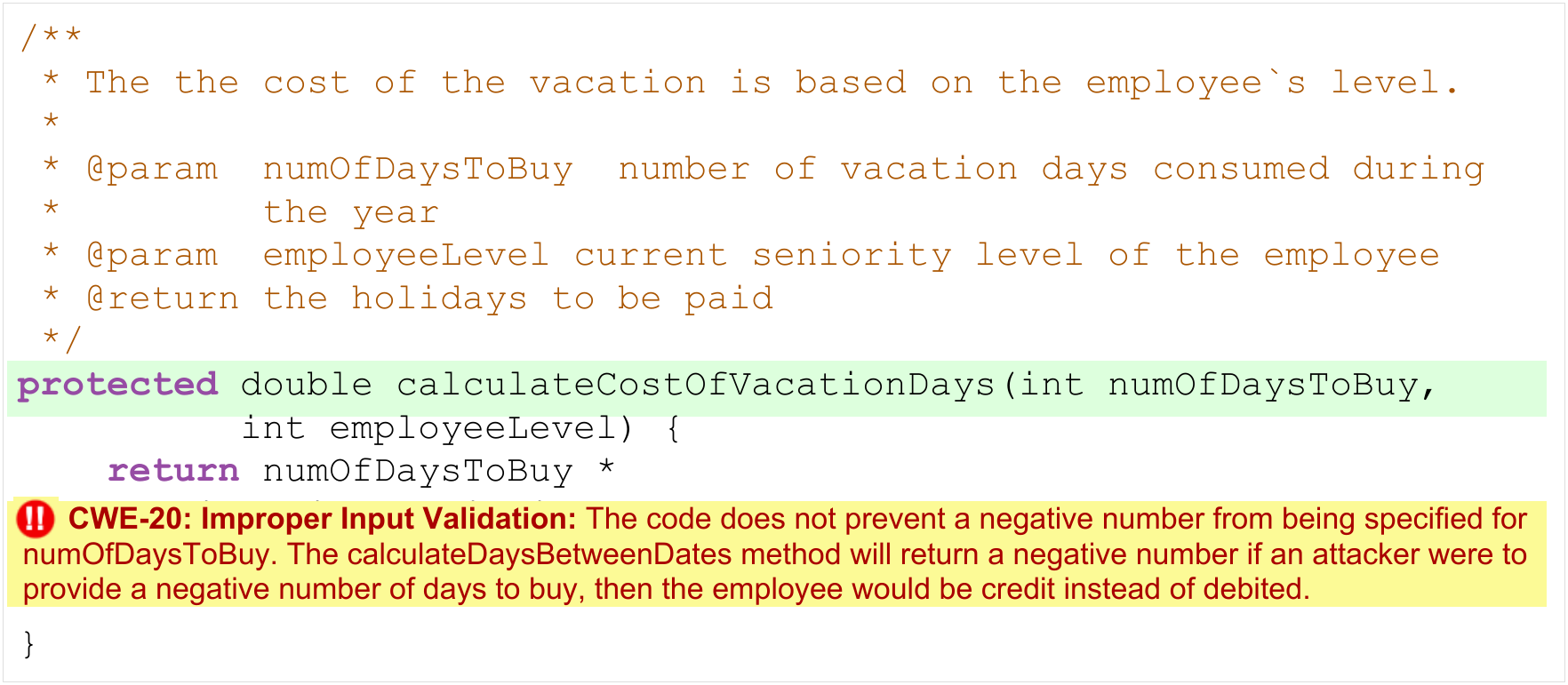}%
	}
	
	\caption{Code snippets with the vulnerabilities in our online experiment.}
	\label{fig:vulnerabilities}
\end{figure}

\noindent\textbf{Security Vulnerabilities.}  We introduced \sqliShort in the code change for the treatment group in the code review experiment, whereas we inject \ivqiShort in the code change for the control group. We select \sqliShort to test our hypothesis because it has a stereotyped attack scenario and is presented with a clear pattern in textbooks~\cite{Clarke:2012, Galluccio:2020}, and even in popular culture~\cite{xkcd}. We selected \ivqiShort as the vulnerability after several brainstorming sessions among the authors and a final validation with the two aforementioned senior software developers. Even though both \sqliShort and \ivqiShort share improper input validation as their root cause and can be neutralized with a solution based on the same principle, the latter does not present a stereotyped attack scenario that can trigger the reviewer's attention. \Cref{fig:vulnerabilities} shows the \sqliShort and the \ivqiShort used in our online experiment. \smallskip

\label{sec:exp-objects}

\noindent\textbf{Control Bug.} One of the main reasons developers perform code review is to detect functional defects~\cite{Bacchelli:2013}. We introduce a bug in the object patch as a robustness check to analyze participants' interest in the task. In other words, if we measure that most participants who do not find the bug also miss the vulnerability, the actual cause could be that those participants do not put enough effort into doing the task and, thus, considering their data in the analyses would likely lead to biased results. In such a case, we would only consider participants who found the bug. We inject a Corner Case (CC) bug, which is typically checked by developers~\cite{spadini2020primers}, as also confirmed by the Google code review guidelines~\cite{GoogleCR} that explicitly encourage developers to check for this type of bug.

\subsection{Pilot Runs}
\label{sec:pilot}

Once the first version of the online experiment was ready, we conducted pilot runs to (1) verify the absence of technical errors in the online platform, (2) check the ratio with which the participants were able to find the injected vulnerabilities (regardless of the treatment group), (3) verify the understandability of the instructions, survey questions, as well as the user interface, (4) improve the code review tool features to ensure that participants' code review experience is as close as possible to an actual one, (5) verify that the code change does not have any design or implementation issues except for the injected vulnerabilities (either \sqliShort or \ivqiShort) and the CC bug, and (6) gather further qualitative feedback from the participants.

We conducted pilot runs for a total of nine participants. The participants' data and qualitative feedback during the pilot runs were discussed iteratively among the authors every few pilot runs. We continued with our pilot iterations until the required changes were minimal. The participants for pilot runs were recruited through the authors' professional network to ensure they would take the task seriously and provide detailed feedback about their experience. In the final study, we used no data gathered from any of the participants who took part in the pilot runs.

% !TEX root = ../security-developers.tex
 
\subsection{Variables, Measurement Details and Analyses}
\label{sec:metho:variables}
%!TEX root=./security-developers.tex
\begin{table}
    \caption{Variables used in the logistic regression models.}\label{tab:vars}
    \centering
    \begin{tabular}{l|l}
    \hline
    \cellcolor[HTML]{C0C0C0}\textbf{Metric}             & \cellcolor[HTML]{C0C0C0}\textbf{Description} \\ \hline\hline
    \multicolumn{2}{c}{\cellcolor[HTML]{EFEFEF}\textit{Dependent Variables}}             \\ \hline
    \vf (\textbf{RQ$_{1}$})  & \begin{tabular}[c]{@{}l@{}}The participant found the vulnerability in the \\ first code review\end{tabular}               \\\hdashline[0.5pt/2pt]
     \vfl (\textbf{RQ$_{2}$}) & \begin{tabular}[c]{@{}l@{}}The participant found the vulnerability during\\revisit to the code review        \end{tabular}                              \\\hline 
     
    \multicolumn{2}{c}{\cellcolor[HTML]{EFEFEF}\textit{Independent Variables}}             \\ \hline
    \vulntype                   & \begin{tabular}[c]{@{}l@{}} Type of the \iivShort vulnerability in the code \\ change (\sqliShort or \ivqiShort)\end{tabular}       \\\hline\hline

    \multicolumn{2}{c}{\cellcolor[HTML]{EFEFEF}\textit{Control Variables   (Review)}}             \\ \hline
    \bug            & \begin{tabular}[c]{@{}l@{}}  The participant found the functional bug \end{tabular}       \\\hdashline[0.5pt/2pt]
    \inter               & \begin{tabular}[c]{@{}l@{}}For how long the participant was\\interrupted during the  review \end{tabular}      \\\hdashline[0.5pt/2pt]
    \reviewDuration                          & \begin{tabular}[c]{@{}l@{}}Duration of the code review\end{tabular}                                                           \\ \hline 
    \multicolumn{2}{c}{\cellcolor[HTML]{EFEFEF}\textit{Control Variables   (Security Knowledge)}}             \\ \hline
    \familiar                      & \begin{tabular}[c]{@{}l@{}} Familiarity to vulnerab.\end{tabular} \\ \hdashline[0.5pt/2pt]
    \courses                      & \begin{tabular}[c]{@{}l@{}} The participant has participated in security \\ courses and/or training\end{tabular} \\ \hdashline[0.5pt/2pt]
    \update                      & \begin{tabular}[c]{@{}l@{}}  The participant keeps himself/herself up to\\date with security information \end{tabular}       \\\hline  
    \multicolumn{2}{c}{\cellcolor[HTML]{EFEFEF}\textit{Control Variables   (Security Practice)}}             \\ \hline 
    \incidents                   & \begin{tabular}[c]{@{}l@{}} The participant has experience with security \\incidents.\end{tabular}       \\\hdashline[0.5pt/2pt]
    \responsible            & \begin{tabular}[c]{@{}l@{}}  The participant looks for vulnerab. as a\\part of his/her job responsibility \end{tabular}       \\\hdashline[0.5pt/2pt]
    \static                       & \begin{tabular}[c]{@{}l@{}}  How often the participant found static\\analysis tools helpful in finding vulnerabilities \end{tabular}       \\\hdashline[0.5pt/2pt]
    \dynamic                  & \begin{tabular}[c]{@{}l@{}}  How often the participant found dynamic\\analysis tools helpful in finding vulnerabilities \end{tabular}       \\\hdashline[0.5pt/2pt]
    \manual                    & \begin{tabular}[c]{@{}l@{}}  How often the participant found manual\\analysis helpful in finding vulnerabilities \end{tabular}       \\\hdashline[0.5pt/2pt]
     \multirow{2}{*}{\shortstack[l]{\{\design/\code/\\\review\}}}&The participant actively considers vulnerabilities\\ 
    & when \{designing software|coding|reviewing code\}    \\ \hline    
    \multicolumn{2}{c}{\cellcolor[HTML]{EFEFEF}\textit{Control Variables   (Team Culture)}}             \\ \hline
    \multirow{5}{*}{\shortstack[l]{\{\tools|\\\library|\\\CRuse|\\\timepressure\}}}&The extend to which developers in the team \\ 
    & \{use tool to detect vulnerabilities | \\
    & check for vulnerabilities in third party libraries |\\
    & use code review to detect vulnerabilities | \\
    & have time to consider security aspects\}   \\\hline  
      \multicolumn{2}{c}{\cellcolor[HTML]{EFEFEF}\textit{Control Variables (demographics)}}             \\ \hline
    \gender                   & \begin{tabular}[c]{@{}l@{}}Gender of the participant\end{tabular}                                                          \\\hdashline[0.5pt/2pt]
    \loe                         & \begin{tabular}[c]{@{}l@{}}Highest achieved level of education\end{tabular}                                                          \\\hdashline[0.5pt/2pt]
    \emp                       & \begin{tabular}[c]{@{}l@{}}Employment status\end{tabular}                                                          \\\hdashline[0.5pt/2pt]
    \role                        & \begin{tabular}[c]{@{}l@{}}Role of the participant\end{tabular}                                                          \\\hdashline[0.5pt/2pt]
    \oss                         & \begin{tabular}[c]{@{}l@{}}The experience in OSS development \end{tabular}                                                          \\\hdashline[0.5pt/2pt]
    \rp                          & \begin{tabular}[c]{@{}l@{}}How often they perform code review\end{tabular}                                                          \\\hdashline[0.5pt/2pt]
    
    \multirow{4}{*}{\shortstack[l]{\{\pde |\\\je |\re |\\\web |\\\db\}}}&Years of experience \{as professional \\ 
    & developer | in java | in code review | in web \\
    & programming | in database applications\} \\ & \\\hdashline[0.5pt/2pt]
    
     \multirow{2}{*}{\shortstack[l]{\{\designFreq |\\\codingFreq | \crFreq\}}}&How often they \{design software | program |\\ 
    & review code\} \\\hline 
    \end{tabular}
    \vspace{-1.5em}
    \end{table}

\textbf{Analysis of the First Code Review Outcome (RQ$_1$).} 
To answer \textbf{RQ$_{1.2}$}, we build a multiple logistic regression model, similar to the one used by McIntosh et al.~\cite{Mcintosh:2016}. The binary dependent variable of our model $\vf$ indicates whether the participant detects the vulnerability during the first review or not (\Cref{tab:vars}). To compute the value of $\vf$, we do the following: (1) the \authorWhoAnalysedCRanswers author inspects all the remarks participants made during the code review experiment and classifies each remark as detection of the vulnerability or not, then (2) the \authorWhoDidQualityCheckforCRanswers author goes through most of the data together with the \authorWhoAnalysedCRanswers author to discuss the decisions, especially the cases that the \authorWhoAnalysedCRanswers author marks as unclear. These authors take the final decision cross-checking their opinion with the answers participants gave to the corresponding question in Step 6 (\Cref{fig:flow}).

To ensure that the selected logistic regression model is appropriate for the data we collect, 
we (1) reduced the number of variables by removing those with Spearman's correlation higher than 0.5 using the VARCLUS procedure, (2) further tested for multicollinearity computing the Variance Inflation Factors (VIF) and removing all values above 7, thus ending with little or no multicollinearity among the independent variables, and (3) built the models adding the independent variables step-by-step and found that the coefficients remained relatively stable, thus further indicating little interference among the variables.

The independent variable $\vulntype$ (\sqliShort or \ivqiShort) is included in the model to investigate how the visibility of a traditional attack scenario for an \iivShort vulnerability affects its detection. To answer \textbf{RQ$_{1.2}$}, we also need to consider the effect of possible confounding factors related to \textit{security knowledge}, \textit{practice}, and \textit{team culture} on the outcome of the code review experiment (i.e., $\vf$). For this reason, we also include in our model a number of control variables (\Cref{tab:vars}). Values for all these variables (except for the ones that regard the review, such as $\bug$ and $\inter$) are collected through the survey questions in Steps 3 and 7 (\Cref{fig:flow}). Details about interruptions ($\interfirst$ and $\intersecond$) are collected from the participants at the end of the reviews, and the duration of each review is computed from the experiment's log.\smallskip

\noindent\textbf{Analysis of the Review Reconsideration (RQ$_2$).}
We build a second multiple logistic regression model to answer \textbf{RQ$_{2.2}$}. The independent and control variables of the second model are the same as those of the first model we build to answer \textbf{RQ$_{1.2}$}, whereas the dependent variable is $\vfl$ (see Table \ref{tab:vars}). The second model is built using data of participants who did not find the vulnerability during the code review task in Step 2 (\Cref{fig:flow}). We used the same approach as for the regression in RQ$_1$ to ensure that the selected model was appropriate.\smallskip

\noindent\textbf{Analysis of open answers on performance.} 
To analyze the answers that participants gave to the open questions when reflecting on the reason for their performance (Step 6, \Cref{fig:flow}), we used open card sorting~\cite{spencer2009card}. This allowed us to extract emerging themes reported as affecting the detection of an \iivShort vulnerability. 
From the open-text answers, the second author created self-contained units, then sorted them into themes. To ensure the themes’ integrity, the author iteratively sorted the units several times. After review by  the first author and discussions, we reached the final themes.
The discussion helped us evaluate controversial answers, reduce potential bias caused by a wrong interpretation of a participant's comment, and strengthen the confidence in the card sorting process's outcomes. 
The card sorting supported us in triangulating our results and form new hypotheses that we challenged with experimental data (\eg end of~\Cref{sec:results:rq1}). The card sorting output is available in our replication package.

\subsection{Recruiting Participants}

The online study was spread out through practitioners' web forums, IRC communication channels, direct authors' contacts from their professional networks, as well as their social media accounts (e.g., Twitter, Facebook). We did not reveal the actual aim of the experiment. We also introduced a donation-based incentive of 5 USD to a charity per participant with a complete and valid experiment.

% \begin{figure}
%	\centering
% 	\includegraphics[width=0.4\textwidth, angle=0]{figs/guidelines.pdf}
%	\caption{Interview Guidelines}
%	\label{fig:design:interviews-guidelines}
%\end{figure}

% !TEX root = ../security-developers.tex

\section{ Results}
\label{sec:study1}

In this section, we describe how we validated the set of participants and report the study results by research question.

\subsection{Valid Participants}
A total of \numOfTotalAccess people accessed the welcome page of our study's web tool through the provided link. Only 194 people went beyond that page and were considered for the experiment.

From these, we excluded instances in which all study steps were not completed, or the first code review (Step 2) was skipped or skimmed (we checked that at least one remark was entered). We manually analyzed the cases of participants who spent less than one-third of the interquartile range in their code review or more than three. Among these, we detected participants who declared to have not done the task seriously and who said they were interrupted significantly during the code review, so their results could not be completely trusted (from this, we removed 10 participants).
After applying the aforementioned exclusion criteria, we had a total of \numParticipantsExperiment valid participants. 

In total, \numParticipantsSQL valid participants received the code change with the \sqliShort, and \numParticipantsIIV received an \ivqiShort. 
We compared the characteristics of the participants assigned to the two groups and found no statistically significant difference.

\begin{figure}[ht]
	\begin{adjustbox}{varwidth=\columnwidth,fbox,center}
		\includegraphics[width=0.9\columnwidth]{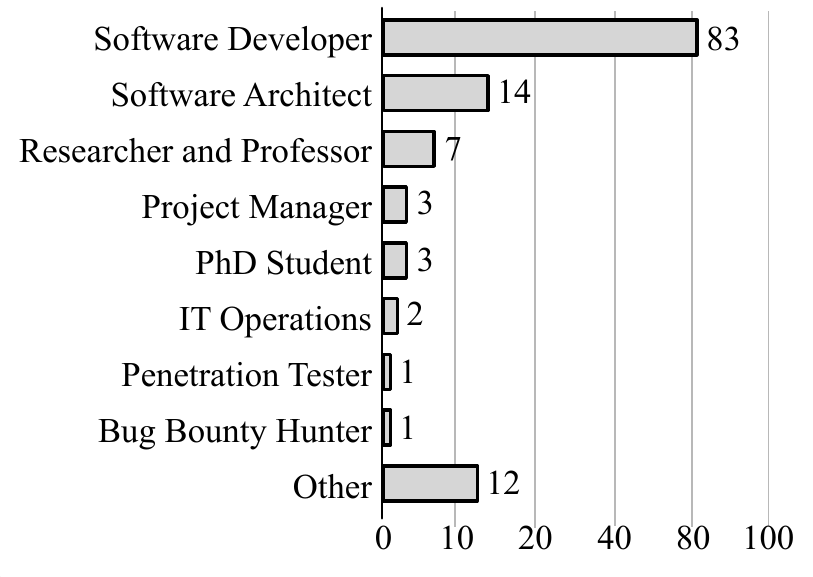}  
	\end{adjustbox}
	\caption{Job of participants with employment.}
	\label{fig:position}
\end{figure}

\begin{figure*}[ht]
	\begin{adjustbox}{varwidth=\textwidth,fbox,center}
		\includegraphics[width=0.45\textwidth]{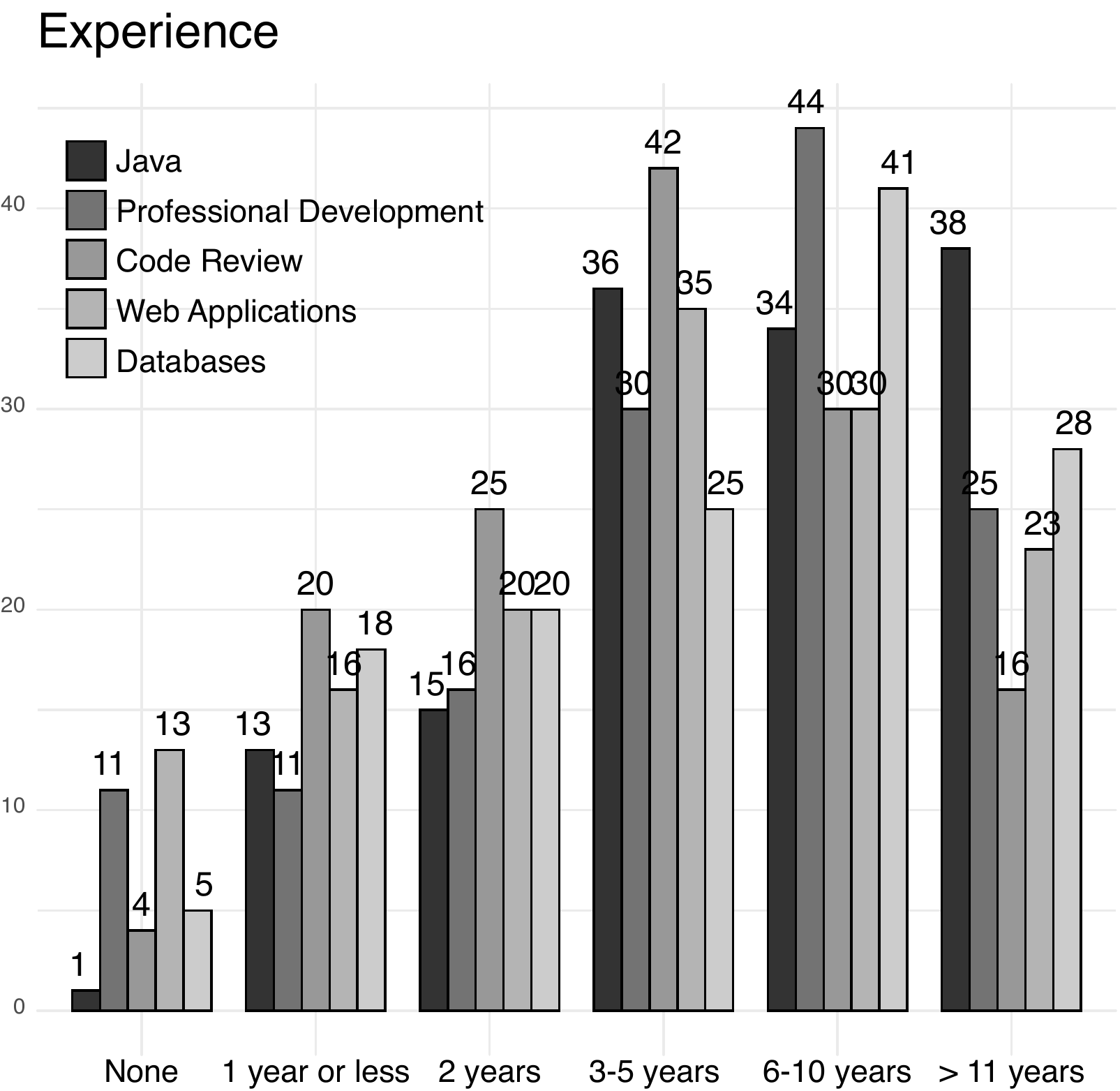}
		\includegraphics[width=0.45\textwidth]{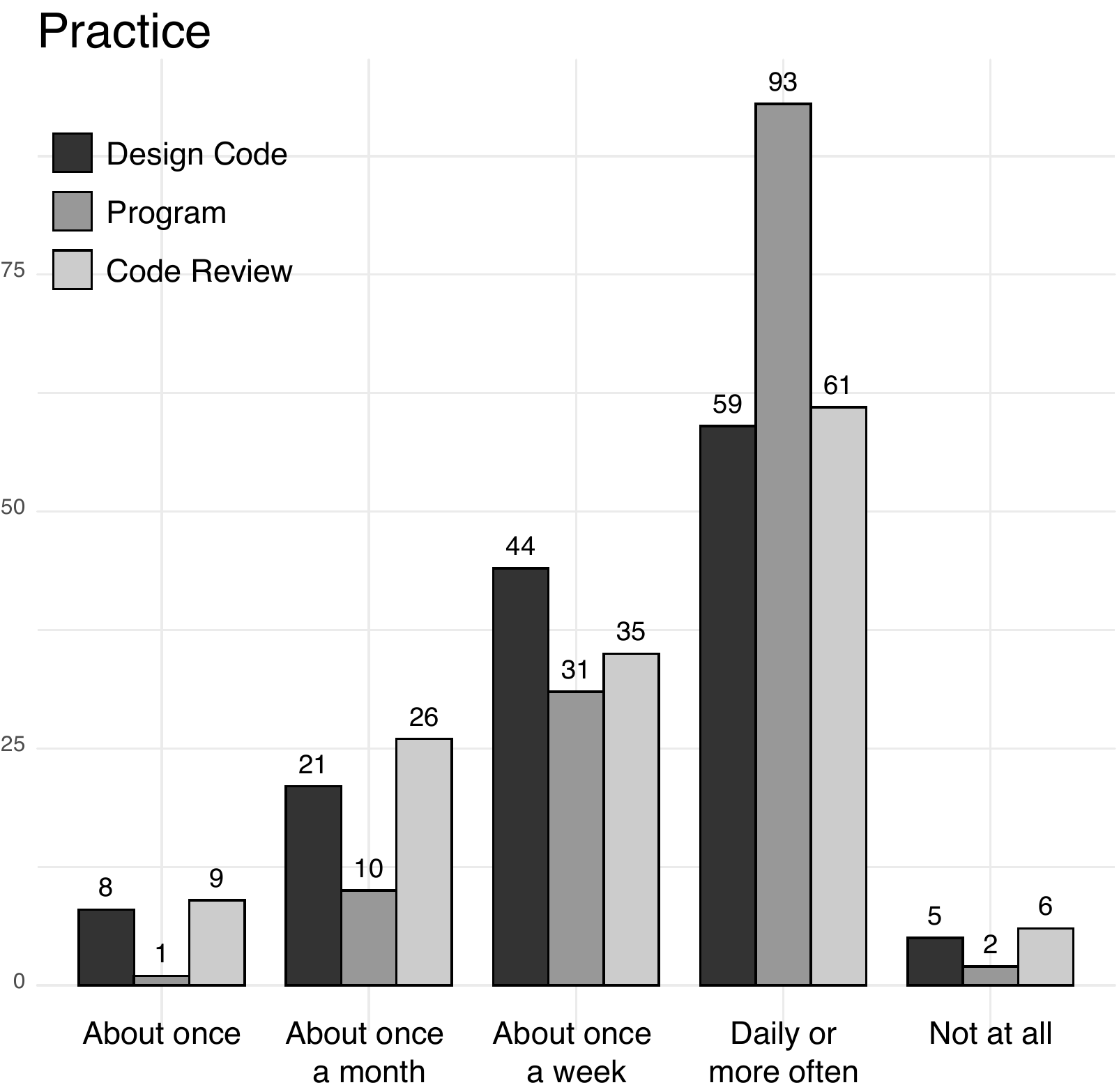}  
	\end{adjustbox}
	\caption{Participants' demographics.}
	\label{fig:demo}
\end{figure*}

In the open-text gender question, \numParticipantsMale and \numParticipantsFemale participants self-described as males and females, respectively, and \numParticipantsGenderNotdisclosed participants preferred not to disclose. The majority of the participants are currently software developers (\perParticipantsSD) and reported to have multiple years of experience in professional software development: \perParticipantsSDthreeyears have 3-5 years of experience, \perParticipantsSDsixyears have 6-10 years, and \perParticipantsSDelevenyears have more than 11 years. Most respondents design, program, and review code daily (\perParticipantsSDdesignDaily, \perParticipantsSDprogramDaily, and \perParticipantsSDcrDaily, respectively). \Cref{fig:position} shows the current positions of participants with part-/full-time employment and \Cref{fig:demo} presents the participants' experience and practice. 

\subsection{RQ$_1$. Detecting  \iivShort vulnerabilities during code review}
\label{sec:results:rq1}
To investigate our first research question, we asked participants to review a code change containing a vulnerability (either an \sqliShort or an \ivqiShort) and a CC bug. 

A total of \numParticipantsFoundBug participants found the CC bug during the first code review, while none of them reported it during the review reconsideration. Among the participants assigned to the code containing the \sqliShort, \numParticipantsSQLFoundBug (\perParticipantsSQLFoundBug) found the CC bug. Among those assigned to the code with the \ivqiShort, \numParticipantsIIVFoundBug (\perParticipantsIIVFoundBug) found the bug. The difference between these groups is not statistically significant, ${\chi}^2 (1,N=146) = 0.77, \pvalue=0.379$.

\begin{table}[ht]
\centering 
 \caption{Detection of the vulnerability in the first review (Step 2).} 
 \label{tab:rq1-oddsratio}
\begin{tabular}{lrrrr}
\multicolumn{1}{l|}{ \iivShort} & \multicolumn{1}{c}{\cellcolor[HTML]{C0C0C0}\textbf{\sqliShort}} & \multicolumn{1}{c}{\cellcolor[HTML]{C0C0C0}\textbf{\ivqiShort}} & \multicolumn{1}{|c}{\cellcolor[HTML]{C0C0C0}\textbf{Total}} \\ 
  \hline
\multicolumn{1}{l|}{Found} & 52 & 14 & \multicolumn{1}{|r}{66} \\ 
  \multicolumn{1}{l|}{Not Found} & 28 & 52 & \multicolumn{1}{|r}{80} \\ 
   \hline
\multicolumn{4}{r}{Odds Ratio: \oddsRatio} \\ 
 \multicolumn{4}{r}{\bf\textit{p} \rhofirst} \\ 
 \multicolumn{4}{r}{~} \\ 
 \end{tabular}
\end{table}

\Cref{tab:rq1-oddsratio} presents the results of the code review task (Step 2 in Figure \ref{fig:flow}) by vulnerability type (\sqliShort vs. \ivqiShort). During this step, 
a total of \numParticipantsFoundFirstCR participants found the vulnerability to which they were assigned. 
Nevertheless, this number is unbalanced.  
Out of the \numParticipantsSQL participants assigned to \sqliShort, \perParticipantsSQLfoundFirstCR found the vulnerability during the review task (Step 2, Figure \ref{fig:flow}).
On the other hand, \numParticipantsIIV participants were assigned to \ivqiShort, and \perParticipantsIIVfoundFirstCR found the \iivShort in this step.
In this review, \numParticipantsIIVmissed participants found neither \sqliShort nor \ivqiShort.
Expressed in odds ratio, these results show how \sqliShort is \oddsRatioshortTXT times more likely to be found by participants than \ivqiShort ($\pvalue$ \rhofirst).

\Cref{tab:regression1} shows the result of the logistic regression model. 
This statistical model confirms the result shown in \Cref{tab:rq1-oddsratio}: Vulnerability type is significant; thus, \emph{we can reject \textbf{H0$_1$}}.

The other statistically significant variables are
\begin{enumerate*} [label=(\roman*)]
	\item $Coding$ (the developer actively considering vulnerabilities when coding) and
	\item $CRFreq$ (how often they review code in the last year).
\end{enumerate*}
All three significant variables have positive estimate values, which means the higher the values of these variables, the more likely the vulnerability is found. 

\begin{table}[ht]
\centering 
 \caption{Regression for the first code review (Step 2).} \label{tab:regression1}
\begin{tabular}{lrrc}
 & \multicolumn{1}{c}{\cellcolor[HTML]{C0C0C0}\textbf{Estimate}} & \multicolumn{1}{c}{\cellcolor[HTML]{C0C0C0}\textbf{S.E.}} & \multicolumn{1}{c}{\cellcolor[HTML]{C0C0C0}\textbf{Sig.}} \\ 
  \hline
Intercept & -12.814 & 3.324 & *** \\ 
  VulnType & 2.351 & 0.574 & *** \\ 
  BugFound & 0.325 & 0.565 &  \\ 
  Interruptions & -0.104 & 0.244 &  \\ 
  Familiarity & 0.954 & 1.247 &  \\ 
  Courses & 0.242 & 0.604 &  \\ 
  KnowledgeUpdate & 0.124 & 0.404 &  \\ 
  Incidents & -0.164 & 0.703 &  \\ 
  Responsibility & -0.388 & 0.323 &  \\ 
  ManualAnalysis & -0.560 & 0.317 &  \\ 
  Coding & 1.079 & 0.422 & * \\ 
  Reviewing & 0.625 & 0.437 &  \\ 
  ThirdPartyLib & 0.108 & 0.266 &  \\ 
  CRusage & -0.372 & 0.344 &  \\ 
  Role & 0.119 & 0.119 &  \\ 
  OSSDev & 0.104 & 0.691 &  \\ 
  DBDevExp & 0.168 & 0.271 &  \\ 
  DevFreq & 0.374 & 0.465 &  \\ 
  CRFreq & 0.963 & 0.418 & * \\
   ... ($\dagger$)       &          &          & \\
   \hline
\multicolumn{4}{r}{Sig. codes:  ‘***’ $p <$ 0.001, ‘**’ $p <$ 0.01, ‘*’ $p <$ 0.05} \\ 
\multicolumn{4}{r}{{\fontsize{6}{6}\emph{($\dagger$) $DurationReview$, $StaticAnalysis$, $ToolUsage$, $EnoughTime$, }}} \\
\multicolumn{4}{c}{{\fontsize{6}{6}\emph{$ProfDevExp$ and $JavaExp$ are not significant and omitted for space reason}}} \\
 \end{tabular}
\end{table}

By analyzing the answers participants gave on why they identified the \iivShort in the first code review, we find recurring themes. In the case of \sqliShort, the top-three reported reasons are:
\begin{enumerate*} [label=(\roman*)]
	\item it is a common and easy to notice vulnerability (18 mentions);
	\item participants have experience with this type of vulnerability (13 mentions); and
	\item participants have knowledge about it (9 mentions).
\end{enumerate*}
These results reflect our expectations: \sqliShort is a traditional attack scenario and, therefore, easier to be recognized. Participants also reported possessing knowledge and previous experience with \sqliShort: \eg having exploited the vulnerability. For instance, a participant reported: ``I know about SQL-injection, and it is still very high on the OWASP Top 10 list, thus whenever I see an SQL statement, I consider the possibility of injection;'' while another stated: ``This is literally school example of SQL Injection with hardcoded SQL and concating parameter.''
However, \sqliShort still frequently happens in practice. For instance, a participant explained: ``SQL Injection issues are among the most common and glaring issues in code that I review.''

For \ivqiShort, developers reported that they follow review practices (10 mentions) to detect this issue: \eg checking user inputs. 
A participant reported the following reason: ``I always consider function inputs and outputs, especially if user-provided.''
Moreover, the reasons reported by the participants show that they do not detect \sqliShort and \ivqiShort the same way, while their root cause is the same, which supports our aforementioned findings.

We challenged these qualitative reasons using data collected in Steps 3 and 7. We used the variables described in \Cref{sec:metho:variables} to map the reasons. All the $knowledge$ variables ($Familiarity$, $Courses$, and $KnowledgeUpdate$) are correlated with participants finding vulnerabilities of this type. We used \textit{Chi-Square} test for the first two and \textit{Mann-Whitney U} test for the last and obtained $\pvalue=0.03$, $\pvalue=0.03$, and $\pvalue=0.01$, respectively. Regarding the participants assigned to \ivqiShort, we performed \textit{Mann-Whitney U} test on the $practice$ variables and only $Dynamic$ was significantly related ($\pvalue=0.02$).

\roundedbox{Developers are \oddsRatioshortTXT times more likely to detect an \iivShort when a traditional attack scenario is visible (\sqliShort) than when it is not (\ivqiShort). Other contributing factors are related to knowledge and practice.}

\subsection{RQ$_2$. Detecting \iivShort after being warned}
To investigate the second research question, we warned the participants about a vulnerability in the code change and invited them to reconsider their review (Step 5 in \Cref{fig:flow}) if they had not found it yet. In this step, no participant found the CC bug. 
\Cref{tab:rq2-oddsratio} presents the results for the reconsidered review.
In total, \numParticipantsFoundSecondCR additional participants identified the vulnerability to which they were assigned. 

We performed a McNemar's test to investigate the effect of warning the participants about a vulnerability in the code change. We considered the code review experiment output for the test. As output, we found that the probability of success is 0.25 with a $\pvalue = 2.152e^{-05}$; considering both \sqliShort and \ivqiShort, the probabilities of success for \sqliShort and \ivqiShort are 0.46 and 0.13 with $\pvalue<0.001$ and $\pvalue=0.02$, respectively. Therefore, \emph{we can reject \textbf{H0$_2$}}. Prompting affects the detection of \iivShort.\smallskip

In total, \numParticipantsSQLfoundSecondCR participants found the \sqliShort. This means that \perParticipantsSQLfoundSecondCRHadMissedFirstCR of the participants who did not find this vulnerability in the first code review found it after the warning. Moreover, \numParticipantsIIVfoundSecondCR participants found the \ivqiShort. 
When expressed in odds, these results show that---when the developers are informed about the existence of a vulnerability in the code (\ie prompted)---the \sqliShort vulnerability is \oddsRatioshortSecondTXT times more likely (\pvalue \rhosecond, \Cref{tab:rq2-oddsratio}) to be found than \ivqiShort.
This result is in-line with that of our first research question, where we also identified that developers are more likely to detect \sqliShort.

\begin{table}[ht]
\centering 
 \caption{Odds ratio for detecting the vulnerability in the review reconsideration (Step 5).} 
 \label{tab:rq2-oddsratio}
\begin{tabular}{lrrrr}
  \multicolumn{1}{c|}{\iivShort} & \multicolumn{1}{c}{\cellcolor[HTML]{C0C0C0}\textbf{\sqliShort}} & \multicolumn{1}{c}{\cellcolor[HTML]{C0C0C0}\textbf{\ivqiShort}} & \multicolumn{1}{|c}{\cellcolor[HTML]{C0C0C0}\textbf{Total}} \\ 
  \hline
 \multicolumn{1}{l|}{Found} & 13 & 7 &  \multicolumn{1}{|r}{20} \\ 
   \multicolumn{1}{l|}{Not Found} & 15 & 45 &  \multicolumn{1}{|r}{60} \\ 
   \hline
\multicolumn{4}{r}{Odds Ratio: \oddsRatioSecond} \\ 
 \multicolumn{4}{r}{\bf\textit{p} \rhosecond} \\ 
 \multicolumn{4}{r}{~} \\ 
 \end{tabular}
\end{table}

In \Cref{tab:regression2}, we show the result of our second logistic regression model. We built the model taking into account only the data of the participants who missed the vulnerability during the first code review. The starting variables used for this statistical model are the same as those used for the first one (see \Cref{sec:metho:variables}), but the final ones differ due to multicollinearity analysis. This model confirms the result shown in \Cref{tab:rq2-oddsratio}: the vulnerability type significantly affects on its detection during the review reconsideration.

\begin{table}[ht]
\centering 
 \caption{Regression for the reconsidered review (Step 5).} \label{tab:regression2}
\begin{tabular}{lrrc}
 & \multicolumn{1}{c}{\cellcolor[HTML]{C0C0C0}\textbf{Estimate}} & \multicolumn{1}{c}{\cellcolor[HTML]{C0C0C0}\textbf{S.E.}} & \multicolumn{1}{c}{\cellcolor[HTML]{C0C0C0}\textbf{Sig.}} \\ 
  \hline
Intercept & -15.261 & 7.427 & * \\ 
  VulnType & 5.584 & 1.770 & ** \\ 
  BugFound & 1.164 & 1.330 &  \\ 
  DurationRevisit & 0.100 & 0.062 &  \\ 
  Interruptions & -0.409 & 0.614 &  \\ 
  Familiarity & 1.934 & 1.905 &  \\ 
  Incidents & -0.186 & 1.572 &  \\ 
  Courses & -0.477 & 1.476 &  \\ 
  KnowledgeUpdate & 1.319 & 0.764 &  \\ 
  Responsibility & -0.690 & 0.624 &  \\ 
  Coding & 0.633 & 0.746 &  \\ 
  Reviewing & 0.735 & 0.997 &  \\ 
  ManualAnalysis & -0.305 & 0.579 &  \\ 
  ToolUsage & -0.144 & 0.637 &  \\ 
  CRusage & 0.355 & 0.636 &  \\ 
  ThirdPartyLib & 0.274 & 0.508 &  \\ 
  EnoughTime & -0.553 & 0.540 &  \\ 
  OSSDev & 0.485 & 1.569 &  \\ 
  ProfDevExp & 0.272 & 0.482 &  \\ 
  JavaExp & 0.482 & 0.640 &  \\ 
  DevFreq & 0.742 & 0.806 &  \\ 
  CRFreq & -0.648 & 0.606 &  \\ 
   ... ($\dagger$)       &          &          & \\
\hline
\multicolumn{4}{r}{Sig. codes:  ‘***’ $p <$ 0.001, ‘**’ $p <$ 0.01, ‘*’ $p <$ 0.05} \\ 
\multicolumn{4}{r}{{\fontsize{6}{6}\emph{($\dagger$) $StaticAnalysis$ is not significant and omitted for space reason}}} \\
 \end{tabular}
\end{table}

Regarding \sqliShort, participants reported that they found the vulnerability in the review reconsideration for the following reasons:  
\begin{enumerate*}  [label=(\roman*)]
	\item they needed to be reminded to focus on security (13 mentions) or
	\item they lacked the confidence (3 mentions).
\end{enumerate*}
Some participants reported more than a reason in their answer.
The first reason is related as participants needed to pay more attention to the security aspect of the code to identify vulnerabilities; as one participant put it: ``At first I was looking mainly for good programming practices and didn't really mind for other issues.'' The last reason refers to participants who found the vulnerability in the first code review, but were initially not confident if it was indeed an issue; later, the warning clarified their doubt. 

Regarding the participants assigned to \ivqiShort, they reported similar reasons:
\begin{enumerate*}  [label=(\roman*)]
	\item they were looking for non-vulnerability defects (3 mentions) or
	\item they needed to be warned about vulnerabilities (2 mentions).
\end{enumerate*}
Similar to participants assigned to \sqliShort, these reported focusing on security when they missed the vulnerability in the first code review and needed to be reminded about security issues to start to look for them (\eg ``... in the second review I was actively looking for security flaws. In the first review I was more inclined to look for code correctness'').  
We see that most participants may know how to identify the \sqliShort correctly and some to detect the \ivqiShort, but they tend not to focus on security during a review.\smallskip

We challenged the reported reasons using data collected in Steps 3 and 7. For the \sqliShort, we related the participants' reasons with variables from $knowledge$ and $practice$ (see \Cref{sec:metho:variables}). We considered the participants that missed the vulnerability in the first code review (\ie they found it in the review reconsideration or completely missed it). We found no significance for neither the $knowledge$ nor the $practice$ variable categories using  \textit{Chi-Square} and \textit{Mann-Whitney U} tests.

\roundedbox{Prompting has an effect on the detection of \iivShort and most reviewers can detect \sqliShort. Yet, a substantial amount of them cannot detect \ivqiShort even after warned of the existence of a vulnerability.}

A total of \numParticipantsMissed participants (\perParticipantsMissed) missed the vulnerability we introduced in the code change.
Respectively, \numParticipantsIIVmissed and \numParticipantsSQLmissed participants did not detect \ivqiShort and \sqliShort.\smallskip

Regarding \sqliShort, developers reported that they missed this vulnerability even after being prompted because they:
\begin{enumerate*} [label=(\roman*)]
	\item lacked knowledge or experience (7 mentions),  
	\item overlooked the details (2 mentions), or
	\item reviewed different aspects of the code (2 mentions).
\end{enumerate*}
Although \sqliShort is well-known, some participants still did not know it. For example, a participant reported: ``I didn't know that this was a vulnerability.'' They also reported to be looking for other details in the code (\eg ``I did not think of the database operation in a detailed way, a mistake of mine.''). These results highlight the need for better security training, even for basic vulnerabilities, and for improving the development process, so security is also considered, especially during code review.\smallskip

Regarding \ivqiShort, developers reported that they missed this vulnerability because they:
\begin{enumerate*} [label=(\roman*)]
	\item lacked attention (10 mentions),
	\item focused on different aspects of the code (9 mentions),
	\item thought that the CC bug was the vulnerability (8 mentions), and
	\item lacked knowledge or experience (6 mentions).
\end{enumerate*}
The reported reasons support developers' belief in their lack of knowledge or experience to detect these vulnerabilities (13 mentions in total), which is in accordance with what SSEs claim~\cite{gitlab-survey}. This reason may be more frequent if we consider ``found a different problem'' as lack of security knowledge as the developers who reported this reason found the algorithmic bug instead of a security issue.\smallskip

We challenged these reasons using the data collected in Steps 3 and 7 (in \Cref{fig:flow}). For the \sqliShort, we investigated the $knowledge$ and $practice$ variables categories (see\Cref{sec:metho:variables}), and found that the following variables are significant: $KnowledgeUpdate$ ($\pvalue=0.01$), $ManualAnalysis$ ($\pvalue = 0.49$), $CodeReviewing$ ($\pvalue = 0.01$), $Coding$ ($\pvalue = 0.01$), and $Design$ ($\pvalue = 0.02$). Regarding the \ivqiShort, we also investigated the $knowledge$ and $practice$ variables categories. We found that most variables are statistically significant. The not significant ones are: $Familiarity$, $Incidents$, and $Practice$.  Our results indicate that both $knowledge$ and $practice$ may be the cause of missing vulnerabilities.

\roundedbox{Most factors related to low knowledge and practice contribute to missing \iivShort vulnerabilities during code review, even after prompting.}

\subsection{Robustness Testing}

To challenge the validity of our findings, we employ robustness testing~\cite{neumayer:2017}. For this purpose, we test whether the results we obtained by our baseline model hold when we systematically replace the baseline model specification with the following plausible alternatives.\smallskip

\noindent\textbf{The functional defect distracted the participants.} A significant number of participants reported that they missed the vulnerability because they searched or found other defects. To verify this claim, we checked the correlation between finding the algorithmic bug and finding the vulnerability. We performed a \textit{Chi-square} test considering all participants, only participants assigned to \sqliShort, and only participants assigned to \ivqiShort. Our analysis did not achieve a statistically significant value for all three cases. Therefore, we do not have enough evidence to suggest a relationship between finding the algorithmic bug and the vulnerability.\smallskip

\noindent\textbf{Vulnerabilities were too easy or too hard to find.} 
 Choosing the right vulnerability to inject in the code change is fundamental to the validity of our results. If a vulnerability is too easy to find, participants might find the issue regardless of any other influencing factor, even without paying too much attention to the review (on the other hand, if it is too complicated, reviewers might not find any vulnerability and get discouraged to continue). On the one hand, in our study, we aimed to evaluate whether developers can identify a standard textbook vulnerability (\sqliShort). Therefore, we expected this vulnerability to be easy-to-catch. In fact, \perParticipantsSQLfound of the participants found it. On the other hand, \ivqiShort is a simple vulnerability to find and fix, but not so recognizable. We measure that ~40\% of the participants found this kind of vulnerability, thus ruling out the possibility that this vulnerability was either too trivial or too difficult to find.\smallskip
 
\noindent\textbf{Number of participants.} We performed a preliminary power analysis using the software package G$^{\star}$Power~\cite{Faul:2007} to calculate the minimum sample size (\ie number of participants with valid responses) for our study. Our prior analysis revealed that we need a minimum sample size of 143 by using \textit{Two-tail} test with \textit{odds ratio} $= 1.5$, $\alpha = 0.05$, \textit{Power} = $1 - \beta$ $= 0.95$, and $R^2$ $=0.3$. We used a manual distribution. As our number of participants (\numParticipantsExperiment) is bigger than necessary, we believe that they are representative. However, this sample size is valid only for the first logistic regression model that we built to answer the research question \textbf{RQ$_{1.2}$}.
To build the second logistic regression model for the review revisit analysis, we exclude the participants that already found the vulnerability in the code review experiment. Therefore, for this analysis, we reduced our participants' number to \numParticipantsCRsecond. Even though this number is quite large in comparison to many experiments in software engineering~\cite{Baum:2019}, it could have affected the significativity of the multivariate statistics; for this reason, we also conducted other statistical tests to verify the effect of single variables on the expected outcome and reported the results.

% !TEX root = ../security-developers.tex
\section{Threats to Validity}

\noindent\textbf{Construct Validity.} 
The code changes we used in our study are a threat to construct validity.  For mitigation purposes, the first and last authors prepared code changes and injected the vulnerabilities and the corner case bug. The other authors later checked the produced code. 
To ensure that participants saw the complete code change, the online platform showed all code on the same page on reasonably sized screens; moreover, participants had to scroll down to proceed to the experiment's next page. 

A major threat is that online experiments could differ from a real-world scenario. 
We mitigated this issue by (1) re-creating a code change as close as possible to a real one (\eg submitting documentation together with the production code), (2) using an interface that is identical to the popular Code Review tool Gerrit~\cite{Gerrit}, (3) injecting vulnerabilities that are based on the examples in the CWE description of both types (\sqliShort and \ivqiShort), and (4) getting the code change validated by two professional software developers. 

To mitigate \textit{mono-operation bias}~\cite{Cook:1979}, we used more than one variable to measure each construct (\eg security knowledge, practice). Each of these variables (see Table \ref{tab:vars}) correspond to a question in the survey on vulnerabilities (Step 3 in \Cref{fig:flow}). To mitigate \textit{mono-method bias}~\cite{Cook:1979}, we used different measurement techniques: We obtained qualitative results by employing card sorting on participants' feedback about why they missed or found the \iivShort vulnerability during the code review task  (Step 2, \Cref{fig:flow}) and code review reconsideration (Step 5). We triangulated these qualitative findings with statistical analyses of variables that we obtained through participants' answers to questions in the survey on vulnerabilities. Finally, to mitigate the \textit{interaction of different treatments}~\cite{Cook:1979}, we applied each treatment separately as follows: (1) Participants were randomly assigned to one of the treatments: \sqliShort or \ivqiShort. We analyzed only the responses participants gave for the code review task (Steps 2) to test the hypothesis \textbf{H0$_{1}$} (i.e., the effect of the visibility of an attack scenario for an \iivShort vulnerability on its detection). (2) To test the hypothesis \textbf{H0$_{2}$} (i.e., the effect of informing participants about the existence of a vulnerability on the detection of \iivShort vulnerability), we analyzed responses that participants who missed \iivShort vulnerability during the code review task gave for the code review reconsideration.\smallskip

\noindent\textbf{Internal Validity.} We reviewed each participation log to identify participants who did not take the experiment seriously. We removed participants who took less than five minutes to complete the experiment or did not complete it. We also introduced a CC bug as a control in the code change for both treatments (\sqliShort and \ivqiShort), as explained in \Cref{sec:exp-objects}. We also checked whether the control bug distracted the participants from finding the vulnerability by conducting a \textit{Chi-square} Test of Independence for all participants, only participants assigned to \sqliShort, and only participants assigned to \ivqiShort---we did not achieve any significant statistical outcome.

As our experiment was online, we cannot ensure that all participants completed it with the same setup (\eg monitor resolution) and similar environments (\eg noise level, interruptions). However, developers in real life also work with different tools in various environments. To mitigate the threats that interruptions might pose to the validity of our study, we asked participants to inform us about durations of interruptions during the code review task and code review reconsideration (Steps 2 and 5 in Figure~\ref{fig:flow}) if there were any. We included these interruptions' durations in our statistical analyses. In addition, several background factors (\eg age, gender, experience, education) may impact the results. Hence, we collected all such information and investigated how these factors affect the results by conducting statistical tests. 
Furthermore, we designed our experiment as a within-subject study to reduce random noise due to participants' differences and to obtain significant results with fewer participants~\cite{MacKenzie:2013, Charness:2012}.\smallskip

\noindent\textbf{External Validity.}
We invited developers from several countries, organizations, education levels, and backgrounds. Nevertheless, our sample is certainly not representative of all developers. 
Thus, further studies are needed to establish the generalizability of our results.

A replication with different vulnerabilities could lead to similar observations as long as they have a similarly popular attack scenario because its effect was clear-cut.

Our observations might not hold when developers review changes to the software projects they work on as a part of their daily practices since higher stakes increase attentiveness~\cite{Reinhart:2014}. However, some participants mentioned that they never consider vulnerabilities.

Moreover, our results may not be the same if large change-sets or changes that address more than one issue are used in the code review experiment: These are more difficult to review as they increase the reviewer’s cognitive load~\cite{Baum:2019}. Therefore, further studies are necessary to assess the generalizability of our results in these scenarios.

% !TEX root = ../security-developers.tex

\section{Discussion}

In this section, we first present themes that emerged as relevant in our study, then provide a high-level overview of the main contributions of our work to research and practice.

\subsection{Emerging Themes}

\smallskip

\noindent\textbf{Lack of security knowledge.}
Software vulnerabilities, such as \iivShort, may have a strong negative impact on software systems, possibly reaching users and even their personal lives. 
Therefore it seems reasonable to think that developers have the knowledge, training, and practice to make sure vulnerabilities do not reach production systems.
However, security experts believe that less than half of developers can actually detect vulnerabilities~\cite{gitlab-survey}. Previous studies~\cite{Woon:2007, Xie:2011} reported that developers' intention to practice secure coding, general security knowledge, and awareness is the cause of vulnerabilities.

We found that the existence of a visible attack scenario facilitates the detection of \iivShort. Developers struggle to recognize vulnerabilities when such a scenario is not available. Indeed, participants were \oddsRatioshortTXT times more likely to find \sqliShort than \ivqiShort since there is no popular example of \ivqiShort in practice. 
Furthermore, many participants reported the lack of knowledge and practice as one of the main reasons for not identifying the vulnerability. 

In-line with previous findings~\cite{Xie:2011, gitlab-survey}, our results suggest the need to improve developers' security knowledge, but they also call for creating different educational approaches. As attack scenarios seem to be more memorable than generic indications on what should be checked and how, educators may focus more on practical scenarios when teaching security. How to design memorable yet effective and recognizable scenarios is an open research question whose answer can have important practical implications.\smallskip

\noindent\textbf{Security is not developers' prime concern.} Developers reported focusing on other kinds of defects and aspects of the code (\eg code quality) as one of the main reasons for not identifying the vulnerability. Indeed, our findings highlighted that prompting developers in searching for a security issue had a significant effect on vulnerability detection. 
Security awareness during code development and the frequency developers perform code review also play a role in it. In line with previous work~\cite{Woon:2007}, these results raise questions on the effectiveness of the current development process, including coding and reviewing activities. To create a different approach, one may consider incorporating explicit security aspects in development activities, such as checklists for code review. The use of code checklists to support developers has been the object of extensive investigation~\cite{Rong:2012,Mcconnell:2004}. Studies can be designed and carried out to determine how to develop security-oriented checklists that do not overburden the reviewers, yet are effective.\smallskip

\noindent\textbf{Practice makes perfect -- with a mentor.} Participants reported how their experience with security issues (or lack thereof) played a key role in detecting (or missing) the \iivShort in the experiment. This raises the question: how can inexperienced developers be trained to find security issues? Our hypothesis is that code review might serve this purpose well. In fact, previous studies~\cite{Bacchelli:2013, Rigby:2013, Sadowski:2018} reported how practical knowledge transfer is one of the main outcomes of the code review process. Therefore, through code review, junior developers can be guided by a more experienced developer in identifying vulnerabilities in the project code-base, with the benefit of having clear real-world examples and scenarios.
Software projects can consider how to integrate this into their code review process and practices.

\subsection{Contributions to Research and Practice}

Overall, our work fits into the context of a type II~\cite{Gregor:2006} middle-range theory~\cite{Wieringa:2015} as we focused on showing how and why software developers can(not) detect improper input validation vulnerabilities.

In this context, the outcomes of our study contribute with the following main points to secure software engineering research and practice:

\begin{itemize}
	\item Our study contributes to cognitive theories of programmer errors~\cite{Vessey:1989,Ko:2005} and debugging~\cite{Atwood:1978}, as well as inspection process models~\cite{Porter:1998}, by providing evidence on the role of explicit attack scenarios, practical knowledge and mindset, and prompting.
	\item Our findings motivate the need for educational research that facilitates the design and implementation of security training for developers by employing authentic~\cite{Lombardi:2007} and experiential learning~\cite{Felicia:2011} techniques. For instance, our study highlights the importance of concrete attack scenarios, suggesting that vulnerabilities with not so popular scenarios should be further explored in security training.
	\item The observed effect of security warnings indicates how code review can be a fertile ground to use vulnerability detectors. Interdisciplinary investigations involving security as well as HCI (Human Computer Interaction) researchers can be conducted with the aim of devising ways to provide this information effectively.
	\item Our study supports that software professionals, particularly developers, should integrate a security-aware attitude into their practices (rather than delegating~\cite{Xie:2011}) to gain the required skills while working on their code-base and avoid overlooking even simple vulnerabilities, such as \ivqiShort.
\end{itemize}

\section{Conclusions}
\label{sec:conclusions}

In the study we presented in this paper, we investigated to what extent developers can(not) detect \iiv vulnerabilities (\iivShort) and the underlying reasons.
To this aim, we designed and conducted an online study that had \numParticipantsExperiment valid participants. These participants were assigned to changes with one of the following two \iivShort types: \sql (\sqliShort) and \ivqi (\ivqiShort). The former vulnerability presents a visible, popular attack scenario.

Overall, \perParticipantsFoundFirstCR of the participants found the vulnerability. Developers were \oddsRatioshortTXT times more likely to detect the \sqliShort, thus confirming the role of the visible attack scenario. After warning the participants of the existence of a vulnerability in the code they just reviewed, an additional \perParticipantsFoundSecondCR of the respondents able to find the vulnerability they missed. Among the \perParticipantsMissed of the participants who could not identify the \iivShort at all, \perParticipantsIIVmissedRateAllMissed were assigned to \ivqiShort.\smallskip

Importantly, these results indicate a lack of knowledge and practice to identify vulnerabilities among the participants, especially when an attack scenario is not visible. The effect of the security warning provides evidence that a significant portion of developers does not focus on security by default, even during code review, but could be triggered to do so with proper team policies or adequate tooling support.

\section*{Acknowledgment}
The authors would like to thank the anonymous reviewers for their thoughtful and important comments, express gratitude to the \numParticipantsExperiment valid participants in the study, and gratefully acknowledge the support of the Swiss National Science Foundation through the SNF Projects No. PP00P2\_170529 and PZ00P2\_186090.

\newpage

\bibliographystyle{IEEEtran}
\bibliography{biblio.bib}

\end{document}